\pdfoutput=1
\documentclass{article}

\usepackage[preprint]{neurips_2021}

\usepackage[utf8]{inputenc} 
\usepackage[T1]{fontenc}    
\usepackage{hyperref}       
\usepackage{url}            
\usepackage{booktabs}       
\usepackage{amsfonts}       
\usepackage{nicefrac}       
\usepackage{microtype}      
\usepackage{xcolor}         

\usepackage{graphicx}
\usepackage{subfigure}
\usepackage{amsmath}

\title{End-to-End Image Compression \\with Probabilistic Decoding}

\author{%
  Haichuan Ma\textsuperscript{1}, Dong Liu\textsuperscript{2}, Cunhui Dong\textsuperscript{1}, Li Li\textsuperscript{2}, Feng Wu\textsuperscript{2}\\
  University of Science and Technology of China\\
  \texttt{\textsuperscript{1}\{hcma, dongcunh\}@mail.ustc.edu.cn} \\
  \texttt{\textsuperscript{2}\{dongeliu, lil1, fengwu\}@ustc.edu.cn} \\
}

\begin{document}

\maketitle

\begin{abstract}
  Lossy image compression is a many-to-one process, thus one bitstream corresponds to multiple possible original images, especially at low bit rates. However, this nature was seldom considered in previous studies on image compression, which usually chose one possible image as reconstruction, e.g. the one with the maximal a posteriori probability. We propose a learned image compression framework to natively support probabilistic decoding. The compressed bitstream is decoded into a series of parameters that instantiate a pre-chosen distribution; then the distribution is used by the decoder to sample and reconstruct images. The decoder may adopt different sampling strategies and produce diverse reconstructions, among which some have higher signal fidelity and some others have better visual quality. The proposed framework is dependent on a revertible neural network-based transform to convert pixels into coefficients that obey the pre-chosen distribution as much as possible. Our code and models will be made publicly available.
\end{abstract}

\section{Introduction}\label{Introduction}

Lossy image compression achieves much higher compression ratio at the cost of incurring information loss, compared with lossless image compression. Due to the information loss, lossy image compression is a many-to-one process, thus one bitstream corresponds to multiple possible original images, especially at low bit rates. However, this nature was seldom considered in previous studies on image compression, including traditional image compression methods such as JPEG \cite{wallace1992jpeg} and JPEG-2000 \cite{skodras2001jpeg}, and recent end-to-end image compression methods \cite{toderici2015variable,toderici2017full,johnston2018improved,balle2016end,balle2018variational,minnen2018joint}. Given a bitstream, all these methods chose one possible image as reconstruction, e.g. the one with the maximal a posteriori probability.

An image compression method usually serves a variety of applications. For example, a compressed image may be viewed by human, or may be analyzed by different machines. Recently, some works have explored image compression techniques for different targets, such as visual quality \cite{rippel2017real,mentzer2020high,tschannen2018deep,blau2019rethinking} and machine task accuracy \cite{agustsson2019generative,choi2020task,chen2019learning,shi2020reinforced}. Although these works have improved the compression performance under specific metrics, this is at the expense of the compression performance under other metrics. Furthermore, some theoretical works have shown that there is inherently trade-off between distortion and perception \cite{blau2018perception,blau2019rethinking}, as well as classification \cite{liu2019classification}. This indicates that an image compression method that decodes only a single image is unable to meet different requirements of different tasks. This creates the demand for multiple reconstruction-based compression method.

However, we emphasize that just obtaining multiple reconstructions is not enough. To serve different tasks, these multiple reconstructions should have diverse characteristics. Among these diverse reconstructions, some are suitable for one task, and some others are suitable for the other task. To show the difference between ``multiple'' and ``diverse,'' we point out that there are some works proposing to add noise into the decoded features to improve the visual quality of reconstructed images \cite{tschannen2018deep,blau2019rethinking}. By adding noise, they can also generate multiple images. However, the obtained multiple images have similar signal fidelity and visual quality, since they optimize their compression model with certain type of distortion.

An intuitive way to enable diverse reconstruction is to use multiple decoders with a single encoder, and different decoders are trained with different distortion metrics. However, this requires knowing all possible tasks before training the model, which is unrealistic, and will increase the storage overhead for storing multiple decoders.
In this paper, our approach is to estimate the distribution of all possible input images that could generate the same bitstream, and directly sample multiple images from it. In this way, during training we only need to minimize the distance between the estimated and the real distributions under the constraint of bit rate. At inference time, the decoder could take different sampling strategies to make the multiple reconstructions diverse.

However, directly estimating the distribution of all possible input images in pixel domain is usually intractable, since the pixels are highly correlated with each other \cite{oord2016pixel}.
Although generative adversarial network (GAN) \cite{goodfellow2014generative,arjovsky2017wasserstein} successfully models the joint distribution among multiple pixels, the joint distribution is captured by the discriminator and could not be expressed explicitly. This makes it impossible to enable sampling mechanism.
Perhaps a possible approach is to modelling the pixels with a autoregressive model \cite{oord2016pixel,salimans2017pixelcnn++}, but this will introduce huge complexity in inference time. In addition, the interdependence among pixels makes it difficult to develop different sampling strategies.
In order to avoid the intractable problem in pixel domain, our solution is to find a tractable feature domain, and enable diverse reconstruction by estimating the distribution of possible features, reshaping the distribution, sampling multiple features from the reshaped distribution, and then transforming the features into images.
Such a feature domain needs to satisfy the condition that the transform between images and features is reversible. Otherwise, minimizing distribution distance in feature domain is not equivalent to minimizing distribution distance in pixel domain.
As a result, the proposed method relies on a reversible neural network-based transform \cite{dinh2014nice, dinh2016density, kingma2018glow, ma2019iwave, ma2020end} to convert pixels to features.

\begin{figure}
  \centering
  \subfigure[]{\includegraphics[width=2.3cm]{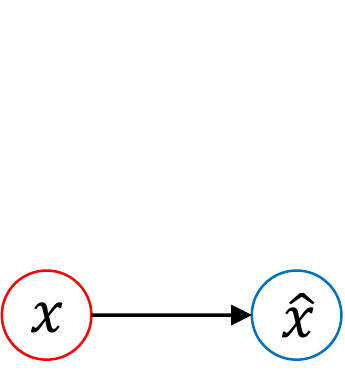}\label{maximal_log_concept}}
  \subfigure[]{\includegraphics[width=2.5cm]{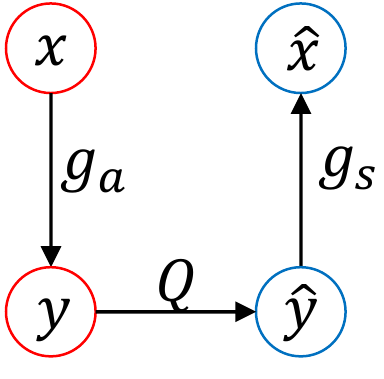}\label{maximal_log_implementation}}
  \subfigure[]{\includegraphics[width=4.0cm]{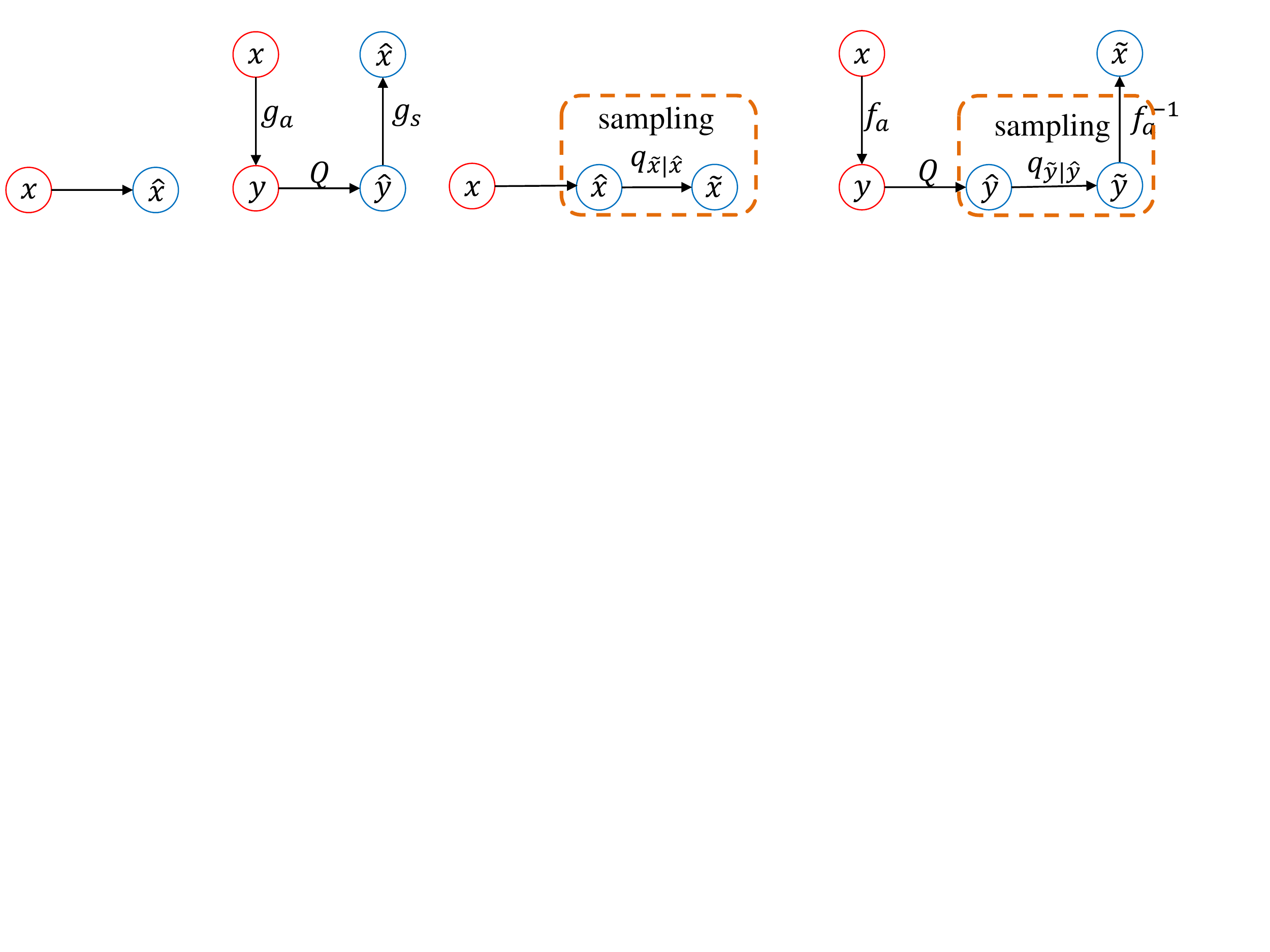}\label{sampling_concept}}
  \subfigure[]{\includegraphics[width=4.2cm]{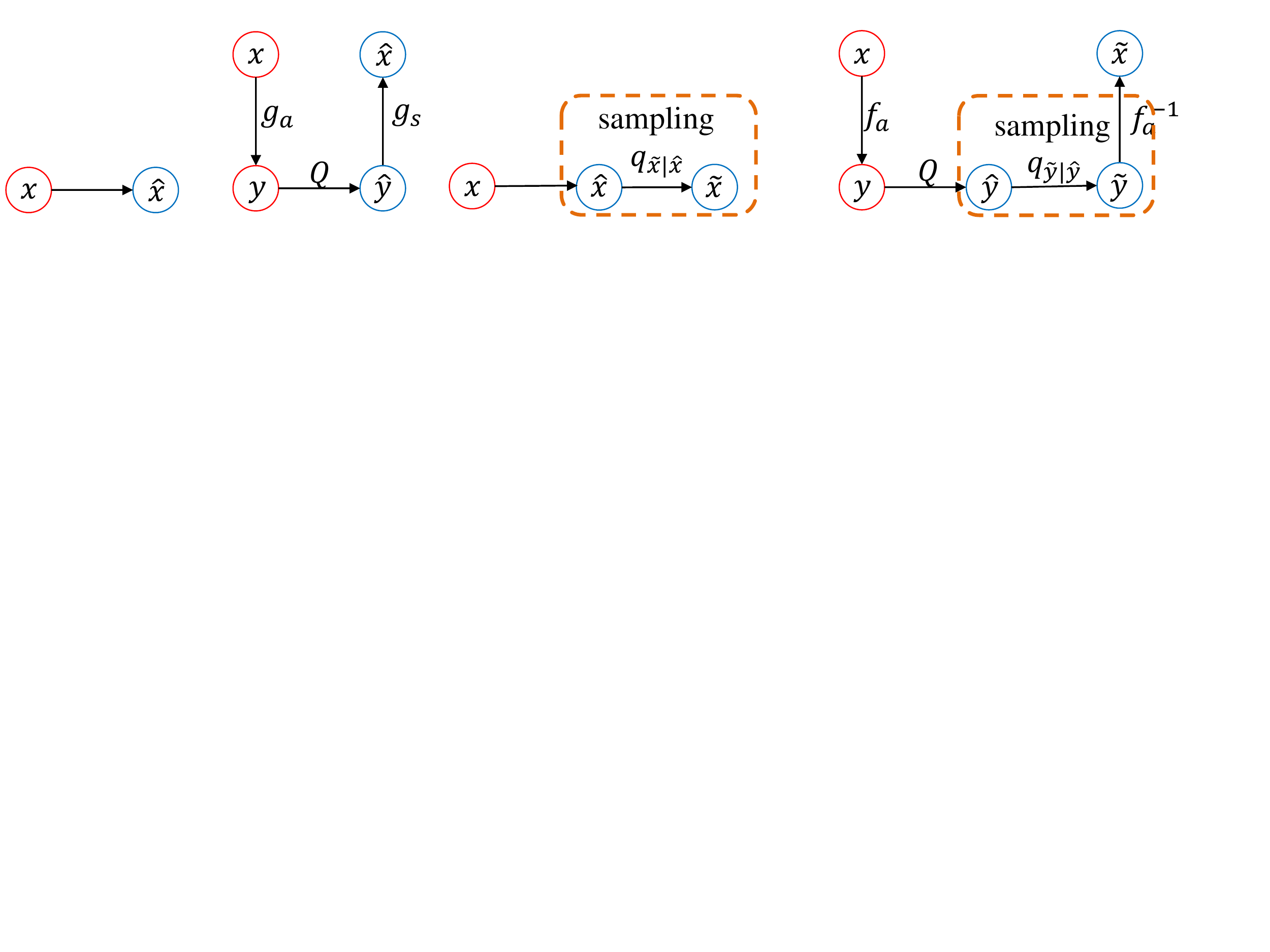}\label{sampling_implementation}}
  \caption{Comparison between deterministic decoding-based image compression ((a) and (b)) and probabilistic decoding-based image compression ((c) and (d)). We omit the same modules between the two to highlight the differences.
  \textbf{(a):} \textbf{Concept} of deterministic decoding-based image compression. For a given bitstream, it outputs one reconstructed image $\hat{x}$. In the reconstructed image, there is inherent tradeoff between different targets, such as signal fidelity, visual quality, and machine task accuracy.
  \textbf{(b):} Popular \textbf{implementation} of deterministic decoding-based image compression via transform coding, where $g_a$ is forward transform and $g_s$ is inverse transform. \textbf{(c):} \textbf{Concept} of probabilistic decoding-based image compression. The decoder could output multiple reconstructed images, by sampling in $q_{\tilde{x}|\hat{x}}$, which is the estimated distribution of all possible input images that can generate the same bitstream. \textbf{(d):} An \textbf{implementation} of probabilistic decoding-based image compression via reversible transform. The decoder firstly samples features in the feature domain, and then transforms the features to pixel domain to reconstruct image. It is much easier to track the distribution $q_{\tilde{y}|\hat{y}}$ of latent features than to track $q_{\tilde{x}|\hat{x}}$ of pixels, because the transform $f_a$ can convert pixels to features that obey the pre-chosen distribution as much as possible. Note that reversible transform makes that sampling in the feature domain is equivalent to sampling in the pixel domain.}
  \label{fig-comparison of compression methods}
\end{figure}

Based on the above intuition, we have made the following contribution in this paper:
\begin{itemize}
    \item For the first time we design a compression framework which allows probabilistic decoding. This is achieved by estimating the distribution of all possible inputs and then sampling multiple outputs from the estimated distribution on the decoder side.
    \item We perform ``first estimating the distribution and then sampling'' in a tractable feature domain instead of the intractable pixel domain. To achieve this, we take advantage of a reversible neural network-based transform to avoid information loss when transforming pixels into features.
    \item We show that by taking different sampling strategies, diverse reconstructions that meet different requirements can be obtained. As a case study, we show that we can get reconstructed images with higher signal fidelity and images with better visual quality, given a single bitstream.
\end{itemize}

\section{Rethinking Image Compression with Deterministic Decoding}\label{Rethinking (End-to-End) Image Compression}

Previous image compression methods take the concept of deterministic decoding, as shown in Fig. \ref{maximal_log_concept}. The popular implementation of this concept is to use transform coding, as shown in Fig. \ref{maximal_log_implementation}. The encoder transforms an image $x$ with a analysis transform $g_a(x;\phi_g)$ into a latent representation $y$, which is then quantized to $\hat{y}$. $\hat{y}$ is written into the bitstream by lossless entropy coding and transmitted to the decoder. The bit rate of the bitstream is calculated by $R(\hat{y})$. On the other side, the decoder recovers $\hat{y}$ from the bitstream, and passes it to synthesis transform $g_s(\hat{y};\theta_g)$ to obtain the output $\hat{x}$.
Previous image compression method is usually trained to minimize the rate-distortion loss,
\begin{equation}\label{RD}
L={\rm E}_{x\sim p_x}\{d(x,\hat{x})+\lambda\cdot R(\hat{y})\}
\end{equation}
where $d(x,\hat{x})$ stands for distortion, and $\lambda$ is used to balance the two terms.

Considering that lossy image compression is a many-to-one process, there is the concept of probabilistic decoding, as shown in Fig. \ref{sampling_concept}. Based on this concept, the optimization problem in (\ref{RD}) can be generalized as follows,
\begin{equation}\label{probabilistic-RD}
L^{\prime} =-{\rm E}_{x\sim p_x}\{{\rm E}_{\tilde{x}\sim p_{\tilde{x}|\hat{x}}}{\rm log}\ q_{\tilde{x}|\hat{x}}(\tilde{x}|\hat{x})+\lambda\cdot{\rm log}\ q_{\hat{y}}(\hat{y})\}
\end{equation}
where $q_{\tilde{x}|\hat{x}}(\tilde{x}|\hat{x})\propto{\rm exp}(-d(\tilde{x}, \hat{x}))$, and $p_{\tilde{x}|\hat{x}}$ is the ground-truth distribution of all possible images $\tilde{x}$ that can generate $\hat{x}$. If $q_{\tilde{x}|\hat{x}}$ is an integrable function, then it can also be regarded as a probability distribution. Then the first term in (\ref{probabilistic-RD}) is the cross entropy between $q_{\tilde{x}|\hat{x}}$ and $p_{\tilde{x}|\hat{x}}$.
Please note that we treat $x$ as one valid sampling from $p_{\tilde{x}|\hat{x}}$, which leads to the distortion term $d(x,\hat{x})$ in (\ref{RD}).

We are now ready to point out the limitations of previous studies on image compression.

\textbf{Limitation 1.} Previous studies usually use pre-defined fixed $d(\cdot,\cdot)$, which makes that $q_{\tilde{x}|\hat{x}}$ is also fixed. However, a fixed $q_{\tilde{x}|\hat{x}}$ prevents (\ref{probabilistic-RD}) from reaching its minimum. This is because that the first term in (\ref{probabilistic-RD}) is the cross entropy between $q_{\tilde{x}|\hat{x}}$ and $p_{\tilde{x}|\hat{x}}$, and when they are equal, this term reaches its minimum. An inappropriate $q_{\tilde{x}|\hat{x}}$ may affect the quality of the reconstructed images. For example, when $q_{\tilde{x}|\hat{x}}$ is i.i.d Gaussian distribution, this is equivalent to using square error as distortion, and the latter has been proven to lead to over-smooth results.

\textbf{Limitation 2.} During the inference time, the reconstructed image is determined as the one with the maximal log-likelihood, i.e.,
\begin{equation}\label{max-log-likelihood}
\hat{x}^{*} = \mathop{\rm argmax}\limits_{\tilde{x}}\ {\rm log}\ q_{\tilde{x}|\hat{x}}(\tilde{x}|\hat{x})
\end{equation}
For example, when $q_{\tilde{x}|\hat{x}}$ is i.i.d. Gaussian distribution, $\hat{x}^{*}= \hat{x}$. This reconstruction manner prevents the decoder from getting diverse reconstructed images.

The above two limitations inspire us to propose a image compression framework that can make full use of probabilistic decoding. This will be described in Section \ref{Probabilistic Image Compression}.

\section{Image Compression with Probabilistic Decoding}\label{Probabilistic Image Compression}
For the convenience of description, we call the proposed image compression with probabilistic decoding as \textit{probabilistic image compression}.
Compared with conventional image compression, probabilistic image compression adopts two technologies to make full use of probabilistic decoding. First, it treats $q_{\tilde{x}|\hat{x}}$ as a trainable model, i.e., (\ref{probabilistic-RD}) becomes
\begin{equation}\label{trainable-probabilistic-RD}
L^{\prime \prime} =-{\rm E}_{x\sim p_x}\{{\rm E}_{\tilde{x}\sim p_{\tilde{x}|\hat{x}}}{\rm log}\ q_{\tilde{x}|\hat{x}}(\tilde{x}|\hat{x};\theta_q)+\lambda\cdot{\rm log}\ q_{\hat{y}}(\hat{y})\}
\end{equation}
where $q_{\tilde{x}|\hat{x}}$ is now a trainable parametric probabilistic function, $\theta_q$ is its trainable parameter set, and others follows the description in Section \ref{Rethinking (End-to-End) Image Compression}. Compared with a stable model, a trainable $q_{\tilde{x}|\hat{x}}$ has potential to converge to $p_{\tilde{x}|\hat{x}}$ automatically during training, if it has strong enough ability to fit any target distribution.

Second, at inference time, the decoder of probabilistic image compression reconstructs image by sampling from the estimated distribution $q_{\tilde{x}|\hat{x}}$, i.e.,
\begin{equation}\label{sampling in pixel}
\tilde{x} \sim q_{\tilde{x}|\hat{x}}
\end{equation}
instead of choosing the one with the maximal likelihood.
Please note that the decoder may adopt different sampling strategies to reconstruct images of different characteristics.

The proposed probabilistic image compression is related to many works, which are summarized below.

\textbf{(1) Relation with lossless image compression.}
Note that in (\ref{trainable-probabilistic-RD}), $\hat{x} = g_s(\hat{y})$ and $g_s$ is a deterministic inverse transform. Then when $\lambda=1$ and $x$ is the only sample from $p_{\tilde{x}|\hat{x}}$, the training objective in (\ref{trainable-probabilistic-RD}) is to minimize the bit rate for losslessly compressing $x$ by transmitting the side information $\hat{y}$.
However, in our framework, only $\hat{y}$ will be written into the bitstream and transmitted, remaining $x$ to be sampled in the decoder side. This is different from lossless compression, which transmits both $\hat{y}$ and $x$.

More essentially, we actually divide the information of $x$ into two parts. The first part is transmitted through bitstream, which is deterministic for the decoder. The other part is uncertainty for the decoder, which is presented in the form of a probability distribution $q_{\tilde{x}|\hat{x}}$. At the same time, we can adjust the value of $\lambda$ in (\ref{trainable-probabilistic-RD}) to adjust the ratio of these two parts of information.

\textbf{(2) Relation with rate-distortion optimization.} Note that the first expected term in (\ref{trainable-probabilistic-RD}) is the cross entropy between $q_{\tilde{x}|\hat{x}}$ and $p_{\tilde{x}|\hat{x}}$. Since the difference between cross entropy and KL-divergency is an unknown constant term ${\rm E}_{\tilde{x}\sim p_{\tilde{x}|\hat{x}}}{\rm log}\ p_{\tilde{x}|\hat{x}}(\tilde{x}|\hat{x})$, the training objective of (\ref{trainable-probabilistic-RD}) is to minimize the KL-divergency between $q_{\tilde{x}|\hat{x}}$ and $p_{\tilde{x}|\hat{x}}$, under the constraint of bit rate of $\hat{y}$.

More importantly, since $q_{\tilde{x}|\hat{x}}$ corresponds to a distortion, i.e. $d(\tilde{x}, \hat{x})\propto-{\rm log}\ q_{\tilde{x}|\hat{x}}(\tilde{x}|\hat{x})$, then using a trainable $q_{\tilde{x}|\hat{x}}$ means that we use a trainable distortion. In our framework, the trainable distortion $d(\cdot,\cdot)$ is determined as the one to minimize the cross entropy between $q_{\tilde{x}|\hat{x}}$ and $p_{\tilde{x}|\hat{x}}$ under certain bit rate after training.

\textbf{(3) Relation with perceptual compression.} Recently, some works introduce GAN into end-to-end image compression to improve the visual quality of compressed images \cite{rippel2017real,mentzer2020high}. In these works, using GAN is to minimize some distance (such as J-S divergency \cite{goodfellow2014generative} or Wasserstein distance \cite{arjovsky2017wasserstein}) between $p_{\hat{x}}$ and $p_x$, which is named as perception. Differently, we choose to minimize the distance between $q_{\tilde{x}|\hat{x}}$ and $p_{\tilde{x}|\hat{x}}$, two conditional distributions conditioned on $\hat{x}$.

Furthermore, as mentioned in Section \ref{Introduction}, since there's tradeoff between distortion and perception, using GAN to train image compression model usually leads to poor signal fidelity. However, in our probabilistic image compression, given a single bitstream, the decoder can generate images with high signal fidelity and images high visual quality at the same time.

\textbf{(4) Relation with deep generative model.} The generative model was originally proposed for modeling data distribution. There are currently four types of generative models that are widely used, including GAN \cite{goodfellow2014generative,arjovsky2017wasserstein}, variational auto-encoder (VAE) \cite{kingma2013auto}, normalized flow model \cite{dinh2014nice, dinh2016density, kingma2018glow}, and autoregressive model \cite{oord2016pixel,salimans2017pixelcnn++}.
The most obvious difference between our method and the above generative models is that they learn the distribution of the whole training set, however, we learn the distribution of the possible images that can generate the same bitstream.
In other word, in probabilistic image compression, we can control the shape of the distribution by transmitting information in the bitstream.
In the special case of zero bit rate, the probabilistic image compression and the deep generative model are the same.

\section{Implementation of Probabilistic Image Compression via Reversible Network}\label{A Realization of Probabilistic Image Compression via Reversible Network}
To implement probabilistic image compression, we need to use a suitable $q_{\tilde{x}|\hat{x}}$ so that it has enough ability to approximate $p_{\tilde{x}|\hat{x}}$. However, estimating the probability distribution of images in pixel domain is intractable. Autoregressive models (such as PixelCNN \cite{oord2016pixel}) are effective, but they introduce huge complexity in inference time. To overcome this, we propose a simple solution that implements probabilistic image compression by taking advantage of reversible neural network. The basic idea is to convert pixels into features without information loss, whose distribution is easy to model. We emphasize that this may not be the only way to implement probabilistic image compression.

\subsection{Overall Structure}\label{Overall Structure}
The block diagram is shown in Fig. \ref{sampling_implementation}.
At the encoder side, the image $x$ is transformed with reversible network-based transform $f_a(x;\phi_f)$ to obtain the latent features $y$, which is then quantized to $\hat{y}$. $\hat{y}$ is written into bitstream by lossless entropy coding. The reversible transform will be described in Section \ref{Revertible Network-based Transform}.

At the decoder side, we first decompress the bitstream to reconstruct $\hat{y}$. Then the synthesis module $g_s(\hat{y}|\theta_g)$ (which is not depicted in the block diagram) generates the parameters $\psi$ of a pre-defined distribution $q_{\tilde{y}|\hat{y}}$, which is the estimated probability distribution of all possible latent features that can be quantized to the same $\hat{y}$. The form of $q_{\tilde{y}|\hat{y}}$ will be given in Section \ref{Form of q}.
Finally, we sample features $\tilde{y}$ from $q_{\tilde{y}|\hat{y}}$, and convert $\tilde{y}$ back to image $\tilde{x}$ with the reversible network-based inverse transform $f_a^{-1}(\tilde{y};\phi_f)$. Note that $f_a^{-1}()$ is the inverse transform of $f_a()$, i.e.,
\begin{equation}\label{inverse-transform}
f_a^{-1}(f_a()) = I
\end{equation}
where $I$ stands for the identity transform.

\subsection{Training Method}\label{Training Method}

The training objective is different from (\ref{trainable-probabilistic-RD}), since we implement probabilistic image compression in the latent feature domain. Our proposed training objective is as follows,
\begin{equation}\label{training objective}
\{\phi_f, \theta_g, \theta_c\} = \mathop{\rm argmax}\limits_{\phi_f, \theta_g, \theta_c}\ {\rm E}_{x\sim p_x}\{{\rm E}_{\tilde{y}\sim p_{\tilde{y}|\hat{y}}}{\rm log}\ q_{\tilde{y}|\hat{y}}(\tilde{y}|\hat{y})+{\rm log\ |det}(\frac{\partial f_a(\tilde{x})}{\partial \tilde{x}})|+\lambda\cdot{\rm log}\ q_{\hat{y}}(\hat{y})\}
\end{equation}
where $p_{\tilde{y}|\hat{y}}$ is the ground-truth distribution of $q_{\tilde{y}|\hat{y}}$.
Note that training model with (\ref{training objective}) is inherently the same as using (\ref{trainable-probabilistic-RD}). This is because $f_a$ is a reversible network-based transform, thus we have
\begin{equation}\label{log-fx-log-x}
{\rm log}\ q_{\tilde{y}|\hat{y}}(\tilde{y}|\hat{y}) = {\rm log}\ q_{\tilde{x}|\hat{x}}(\tilde{x}|\hat{x}) + {\rm log\ |det}(\frac{\partial f_a(\tilde{x})}{\partial \tilde{x}})|
\end{equation}
where $\hat{x}=f_a^{-1}(\hat{y})$.

Although we don't know the real distribution $p_{\tilde{y}|\hat{y}}$, we have a valid sample from it, the direct transformed feature $y$ of the input image $x$, which allows us to train our model. On this condition, the training objective in (\ref{training objective}) can be simplified as follows.
\begin{equation}\label{simplified training objective}
\{\phi_f, \theta_g, \theta_c\} = \mathop{\rm argmax}\limits_{\phi_f, \theta_g, \theta_c}\ {\rm E}_{x\sim p_x}\{{\rm log}\ q_{\tilde{y}|\hat{y}}(y|\hat{y})+{\rm log\ |det}(\frac{\partial f_a(x)}{\partial x})|+\lambda\cdot{\rm log}\ q_{\hat{y}}(\hat{y})\}
\end{equation}

In experiment, we find that only using (\ref{simplified training objective}) to train the model will lead to meaningfulness reconstructed images. This may be because that training with (\ref{simplified training objective}) does not keep low frequency information, and non-linear transform also amplifies the errors in the feature domain.
As a result, we introduce a regularization term, then the final training objective is as follows,
\begin{equation}\label{simplified training objective with regularization}
\{\phi_f, \theta_g, \theta_c\} = \mathop{\rm argmax}\limits_{\phi_f, \theta_g, \theta_c}\ {\rm E}_{x\sim p_x}\{{\rm log}\ q_{\tilde{y}|\hat{y}}(y|\hat{y})+{\rm log\ |det}(\frac{\partial f_a(x)}{\partial x})|-d(x,x^0)+\lambda\cdot{\rm log}\ q_{\hat{y}}(\hat{y})\}
\end{equation}
where $d(\cdot,\cdot)$ is mean square error, $x^0=f_a^{-1}(y^0)$, and $y^0$ is the mode of $q_{\tilde{y}|\hat{y}}$. Note that $d(\cdot,\cdot)$ may take other forms.

\subsection{Revertible Network-based Transform}\label{Revertible Network-based Transform}
We use the same transform module as in \cite{ma2019iwave,ma2020end}, which is named as wavelet-like transform. For a two-dimensional image, we first split it into two parts, \{$L$, $H$\}, which are named as subband, according to odd and even rows. Then we use 2 lifting steps to remove the correlation between them, where one lifting step is formulated as follows,
\begin{equation}
\label{Coupling-f}
\begin{split}
\left\{
\begin{aligned}
&H = H + t_L(L)\\
&L = L + t_H(H)
\end{aligned}
\right.
\end{split}
\end{equation}
where $t_L$ and $t_H$ are all implemented with neural networks. The decorrelated $L$ and $H$ are then decomposed in the column direction and undergo another 2 lifting steps respectively, thus forming the $1_{st}$-level decomposition result, \{$LL$, $HL$, $LH$, $HH$\}. Usually one of the four part, such as $LL$, will be iteratively decomposed to form a pyramid structure.

For the inverse transform, we just need to use the inverse mode of lifting to process every two subbands, and then merge them together. The formulation of inverse mode of lifting is as follows,
\begin{equation}
\label{Coupling-i}
\begin{split}
\left\{
\begin{aligned}
&L = L - t_H(H)\\
&H = H - t_L(L)
\end{aligned}
\right.
\end{split}
\end{equation}
where the neural network-based $t_L$ and $t_H$ share their weights with the corresponding blocks in the forward transform.

Supposing the input image $x$ having the spatial size of $h\times w$, after $K$-level of wavelet-like forward transform, we obtain $3K+1$ subbands of wavelet coefficients, i.e. $y=\{HH_1, LH_1, HL_1, ..., HL_K, LL_K\}$. The $K_{th}$-level subbands ($LL_K$, $HL_K$, $LH_K$, and $HH_K$) have the shape of $\frac{h}{2^K}\times \frac{w}{2^K}$, and the $1_{st}$-level subbands ($HH_1$, $LH_1$, and $HL_1$) have the shape of $\frac{h}{2}\times \frac{w}{2}$.

Note that using wavelet-like transform, the second term in (\ref{simplified training objective with regularization}) is zero.

\subsection{Form of $q_{\tilde{y}|\hat{y}}$}\label{Form of q}
We use a factorized Gaussian model to represent the estimated distribution of all possible input features, i.e.,
\begin{equation}
\label{factorized Gaussian}
q_{\tilde{y}|\hat{y}}(\tilde{y}|\hat{y}) = \prod_{i}N(\tilde{y}_i|\mu_i(\hat{y}),s_i(\hat{y}))
\end{equation}
where $\mu_i()$ and $s_i()$ are neural network-based modules to synthesis the mode and variance of the Gaussian model.

Although we still use coefficient-independent probability distribution, this is essentially different from using pixel-independent probability distribution.
The reason is that the transform module is trainable so that it can convert pixels into coefficients that obey the distribution as possible.

\subsection{Quantization}\label{Quantization}
All the subbands first need to be quantized in order to enable the following lossless entropy coding process. During inference time, we directly quantize $y$ to $\hat{y}$ with rounding operation. However, rounding operation is not suitable during the training process because it will stop the backpropagation process.
In training, we take advantage of the soft-to-hard quantization strategy \cite{agustsson2020universally} to make the gradients pass through the quantization layer.

\subsection{Bit Rate Estimation}\label{Bit Rate Estimation}
In this section, we describe how to calculate $q_{\hat{y}}(\hat{y})$ in (\ref{simplified training objective with regularization}). For every coefficient in every subband, we represent its distribution with 3 weighted Gaussian models, whose modes, variances and weights are estimated with neural network-based context model by analysing its context. For example, for the j-th coefficient in the i-th subband, its estimated distribution is as follows,
\begin{equation}\label{prob}
q^{\prime}_{\hat{y}_{i,j}}(\cdot)= \sum_{n=1}^{3}\omega_{i,n}(c_{i,j})\cdot N(\cdot|\mu_{i,n}(c_{i,j}), s_{i,n}(c_{i,j}))
\end{equation}
where $\omega_{i,n}()$, $\mu_{i,n}()$ and $s_{i,n}()$ are neural network-based modules, and $c_{i,j}$ is the context of the coefficient $\hat{y}_{i,j}$.
Based on (\ref{prob}), $q_{\hat{y}}(\hat{y})$ in (\ref{simplified training objective with regularization}) can be calculated as follows,
\begin{equation}\label{rate}
q_{\hat{y}}(\hat{y})= \prod_{i,j} \int_{\hat{y}_{i,j}-0.5}^{\hat{y}_{i,j}+0.5}q^{\prime}_{\hat{y}_{i,j}}(\hat{y}_{i,j})
\end{equation}

In order to utilize the correlation between subbands to improve the compression efficiency, all subbands need to be compressed jointly. For a coefficient in subband $LL_K$, its context is the upper and left coefficients in its surrounding $s\times s$ area. For a coefficient in the subband $HL_k$, its context is the upper and left coefficients in its surrounding $s\times s$ area, and all the coefficients in the co-located $s\times s$ area of $LL_k$, where
$$LL_k = f_a^{-1}(LL_{k+1}, HL_{k+1}, LH_{k+1}, HH_{k+1})$$
for $k<K$.
For the coefficients in subbands $LH_k$ and $HH_k$, the way of constructing context is similar.

\section{Experiments}\label{Experiments}
\subsection{Training Details}\label{Training Details}
Our probabilistic image compression model can be trained as previous end-to-end image compression methods.
The training set consists of 800 high-definition (HD) images in the DIV2K \cite{agustsson2017ntire} training set. During training, we randomly crop patches of size $128\times128$ from the entire training set. Random horizontal and vertical flipping are used for data augmentation. We train the neural network using Adam \cite{kingma2014adam} optimizer with $\beta_1=0.9$ and $\beta_2=0.999$. The learning rate is set to 1e-4, and the batch size is set to 8. We train our models with 2 NVIDIA GeForce RTX 2080 Ti GPUs.

We first initialize other neural network-based modules except for the transform module by training them with the traditional CDF9/7 wavelet transform \cite{skodras2001jpeg}. Then we replace the CDF9/7 transform with reversible network-based transform, and train the whole model for 100K iterations. The initialization step helps improve the performance and speeds up the joint training process.

\subsection{Learned Distribution $q_{\tilde{y}|\hat{y}}$ of Possible Input Coefficients}\label{Learned Distribution}
\begin{figure*}
  \centering
  \subfigure[Input Image]{\includegraphics[width=3cm]{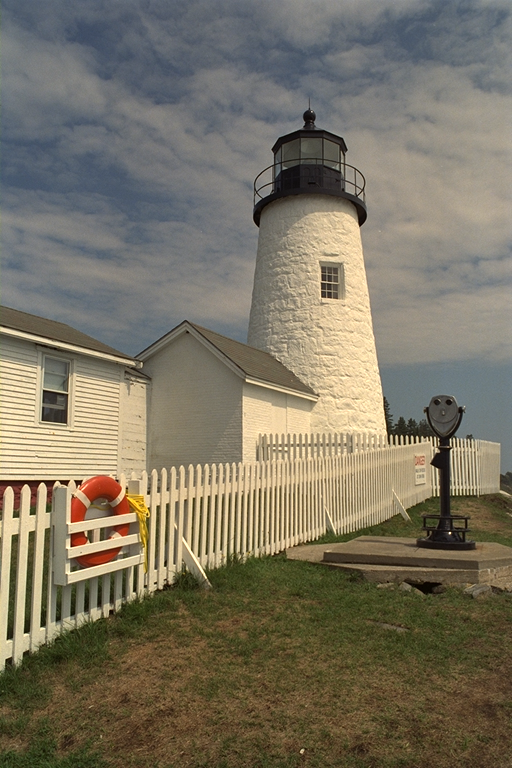}}
  \subfigure[Variances at 0.2 bpp]{\includegraphics[width=3cm]{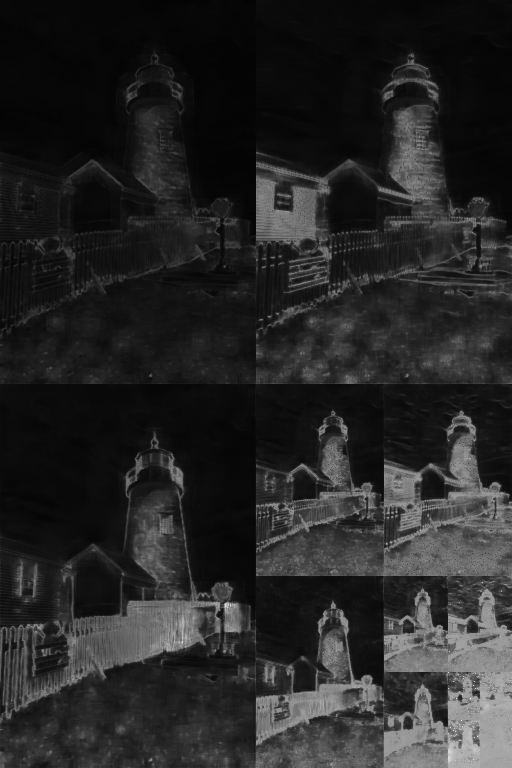}}
  \subfigure[Variances at 0.7 bpp]{\includegraphics[width=3cm]{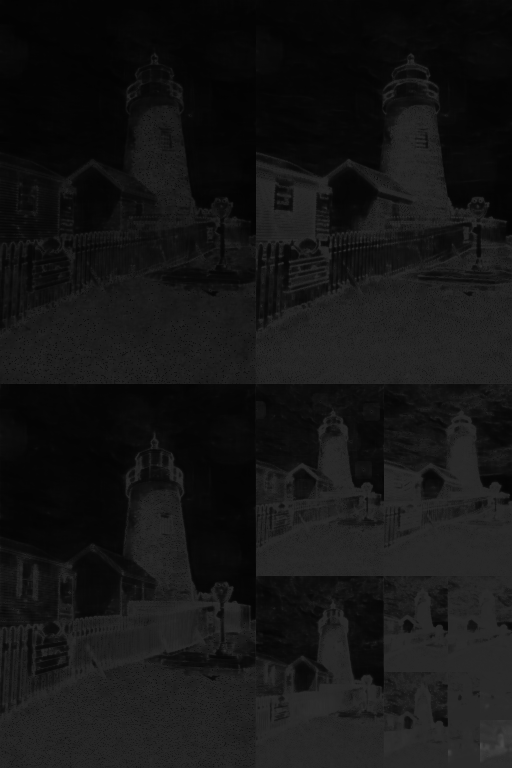}}
  \caption{Visualization of the estimated variances of the learned distribution $q_{\tilde{y}|\hat{y}}$ at different bit rate. In order to make better use of 8-bit depth for display, we have enlarged all the variances by 7 times. It can be seen that bitstream of lower bit rate generates distribution with larger variances.}
  \label{scales of q}
\end{figure*}
The most obvious difference between probabilistic image compression and deterministic decoding-based image compression is that it attempts to model the information lost during the compression process and express it as the variances of $q_{\tilde{y}|\hat{y}}$. To illustrate this point, we visualize the variances of the learned distribution $q_{\tilde {y}|\hat{y}}$ in Fig. \ref{scales of q}. It can be found that the variances are greater at lower bit rate. This is reasonable, because a lower bit rate means that less information is transmitted. Therefore, more information is captured by $q_{\tilde{y}|\hat{y}}$, resulting in greater variances.

On the other hand, Fig. \ref{scales of q} also shows that the variances between different locations of the same subband are different. Considering the relation between the estimated distribution and the distortion, i.e. $d(\tilde{y}, \hat{y})\propto-{\rm log}\ q_{\tilde{y}|\hat{y}}(\tilde{y}|\hat{y})$, this actually means that we use different distortions for different locations.
This shows that during the training process, our method could automatically learn a distortion metric.

\subsection{Diverse Reconstruction with a Single Bitstream}\label{Perception-Distortion Tradeoff with a Single Bitstream}
\begin{figure}[htbp]
\centering
\begin{minipage}{14cm}
\subfigure[LPIPS-MSE]{\includegraphics[width=5.8cm]{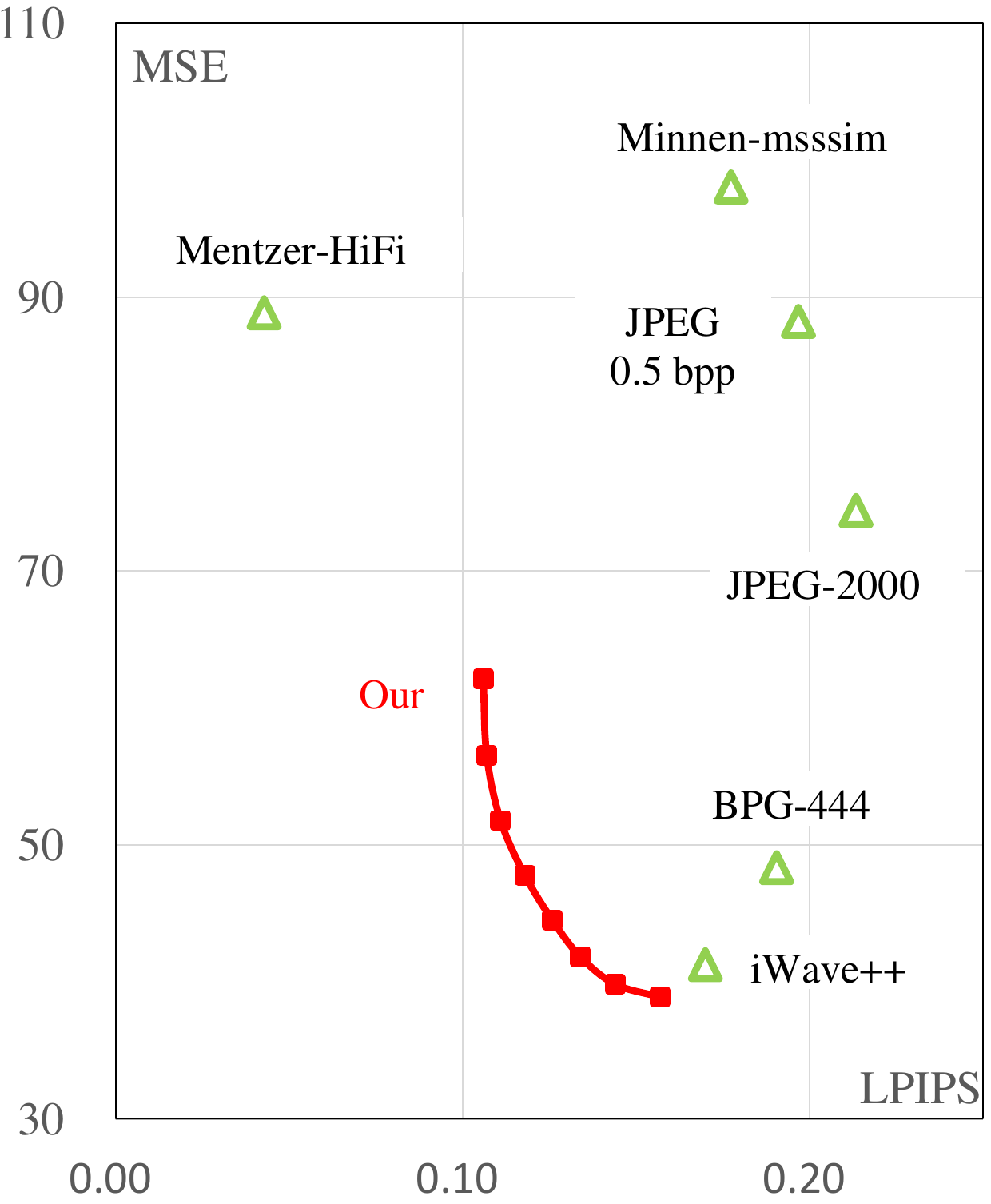}}
\hspace{0.5cm}
\subfigure[Original Image]{\includegraphics[width=5.1cm, height=6.9cm]{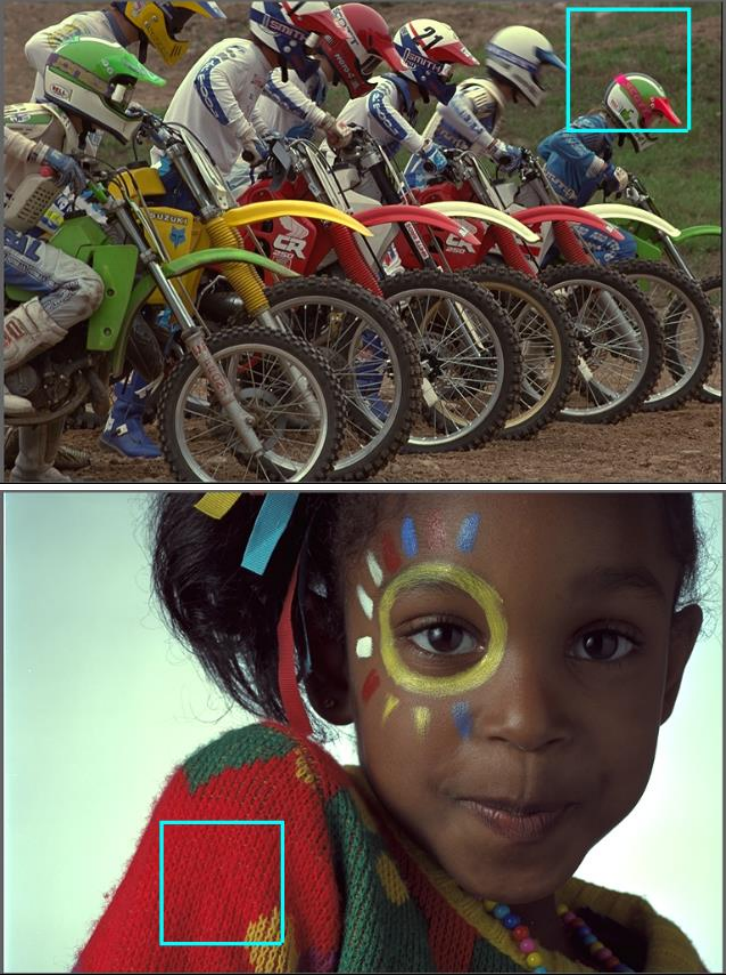}}
\end{minipage}
\begin{minipage}{14cm}
\subfigure{\includegraphics[width=2.5cm]{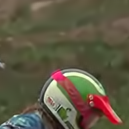}}
\subfigure{\includegraphics[width=2.5cm]{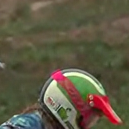}}
\subfigure{\includegraphics[width=2.5cm]{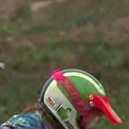}}
\subfigure{\includegraphics[width=2.5cm]{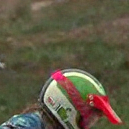}}
\subfigure{\includegraphics[width=2.5cm]{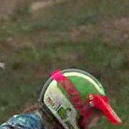}}
\end{minipage}
\vspace{-0.1cm}
\begin{minipage}{14cm}
\subfigure{\includegraphics[width=2.5cm]{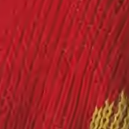}}
\subfigure{\includegraphics[width=2.5cm]{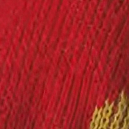}}
\subfigure{\includegraphics[width=2.5cm]{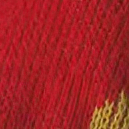}}
\subfigure{\includegraphics[width=2.5cm]{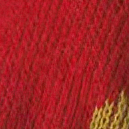}}
\subfigure{\includegraphics[width=2.5cm]{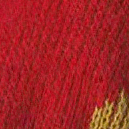}}
\end{minipage}
\caption{\textbf{(a):} Traveling between MSE and LPIPS \cite{zhang2018unreasonable} on Kodak dataset at 0.34 bpp for probabilistic image compression, by changing the variances of $q_{\tilde{y}|\hat{y}}$ at the decoder. Please note that MSE is used to reflect the signal fidelity, and LPIPS is used to reflect the visual quality. The compared methods have similar or higher bit rate than our method, and could not achieve such a traveling. ``Mentzer-HiFi'' refers to \cite{mentzer2020high}, ``Minnen-msssim'' refers to \cite{minnen2018joint}, ``iWave++'' refers to \cite{ma2020end}. \textbf{(b):} Original images. \textbf{(Bottom):} Visual examples of diverse reconstructions of probabilistic image compression. From left to right, the variances of $q_{\tilde{y}|\hat{y}}$ is set to \{0, 0.3, 0.5, 0.7, 1.0\} times of the estimated values, and it can be seen that the visual quality becomes better.}
\label{Fig PD tradeoff}
\end{figure}

The most obvious feature of probabilistic image compression is that it reconstruct images by sampling. Then by taking different sampling strategies, we could obtain reconstructed images with different characteristics. As a simple case, in Fig. \ref{Fig PD tradeoff}, we show that we can control the signal fidelity (measured by MSE) and visual quality (measured by LPIPS \cite{zhang2018unreasonable}) of the reconstructed image. This can be achieved very simply by uniformly adjusting the variances of the estimated distribution. Smaller variances lead to better signal fidelity, and larger variances lead to better visual quality.
We emphasize that exploring other sampling methods may be beneficial for other targets, such as machine tasks.

\subsection{Compression Performance}\label{Overall Compression Performance}

We provide the overall compression performance of our implementation for probabilistic image compression on Kodak dataset. The compared methods are two traditional methods, JPEG-2000 and BPG\footnote{https://bellard.org/bpg/}, and four end-to-end methods, ``Minnen-msssim'' \cite{minnen2018joint}, ``Minnen-mse'' \cite{minnen2018joint}, ``iWave++'' \cite{ma2020end} and ``Mentzer-HiFi'' \cite{mentzer2020high}. Among them, ``Mentzer-HiFi'' is trained for better visual quality by using LPIPS as optimization objective and adversarial training. ``iWave++'' could be regarded as our baseline model, since we implement probabilistic decoding based on it.

Two metrics are used for evaluation, including PSNR, which is used to measure the signal fidelity, and LPIPS, which is used to measure the visual quality. For our method, we calculate PSNR by setting the variances of $q_{\tilde{y}|\hat{y}}$ to zero. When calculating LPIPS, at every bit rate, we find the lowest value of LPIPS by uniformly adjusting the variances.

\begin{figure*}
  \centering
  \subfigure[bpp-PSNR]{\includegraphics[width=5cm]{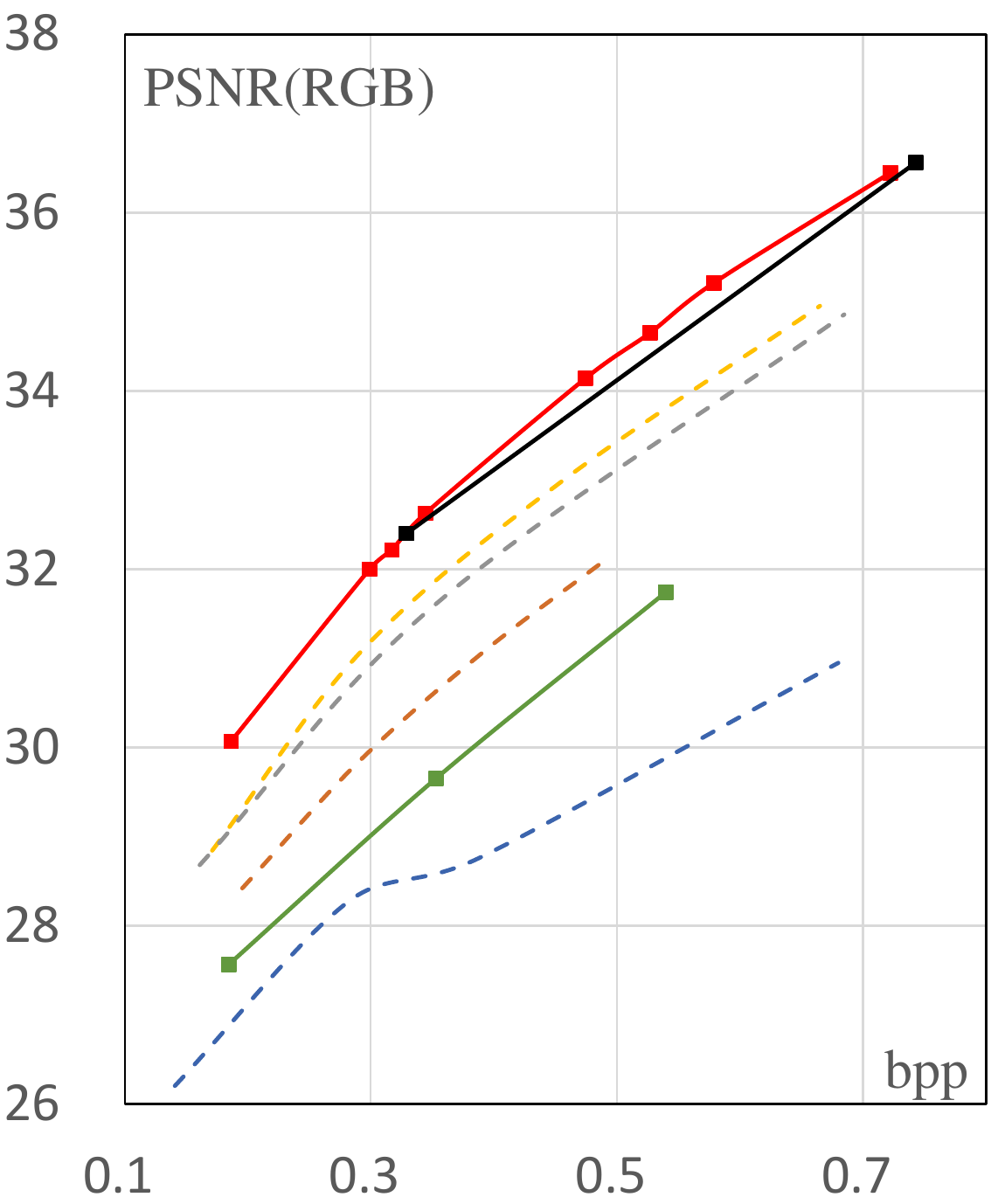}}
  \hspace{0.5cm}
  \subfigure[bpp-LPIPS]{\includegraphics[width=5cm, height=6.0cm]{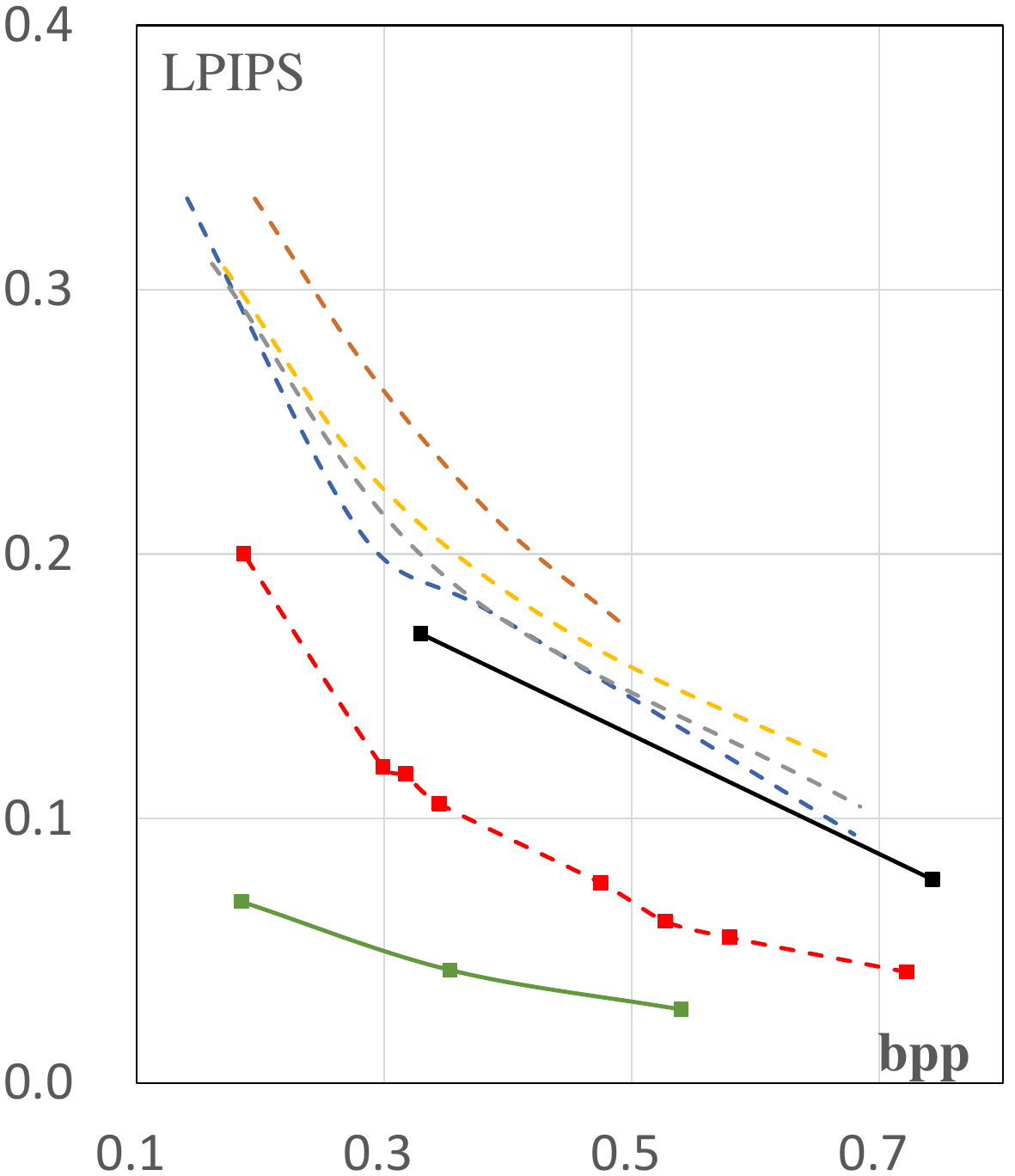}}
  \hspace{0.5cm}
  \subfigure{\includegraphics[width=2.3cm]{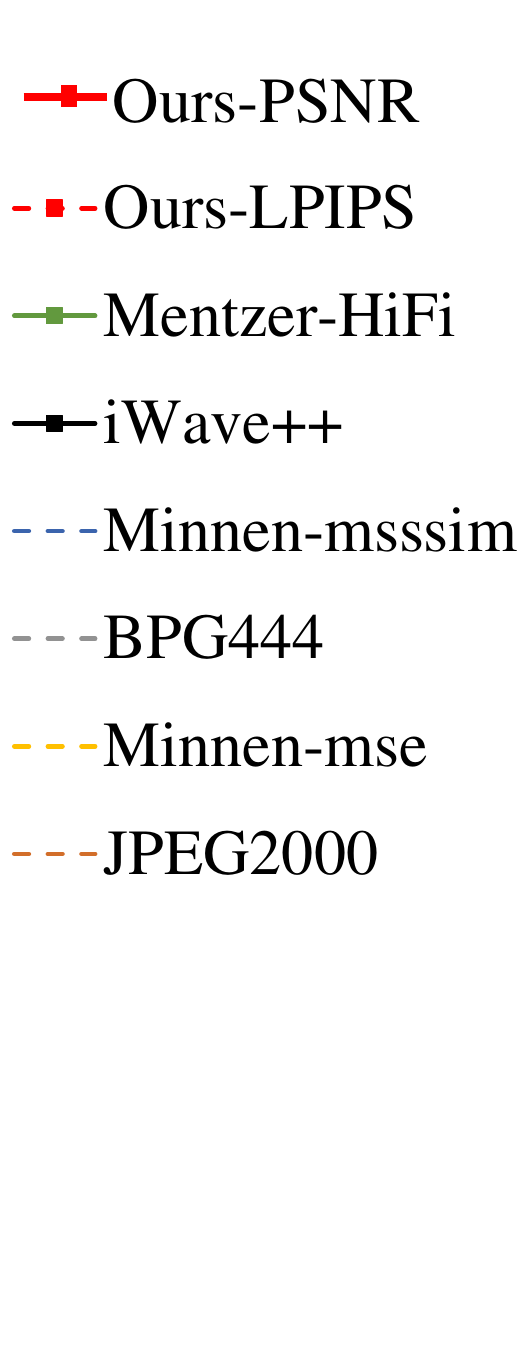}}
  \caption{Rate-distortion curves of different lossy compression methods on the Kodak dataset. Please note that PSNR is used to reflect the signal fidelity, and LPIPS \cite{zhang2018unreasonable} is used to reflect the visual quality. ``Ours-PSNR'' and ``Ours-LPIPS'' are using reconstructed images with best PSNR and best LPIPS, respectively. However, we emphasize the different images are generated with the same bitstream.
  Our method performs well under both metrics, but the compared methods perform poorly under at least one metric. ``Mentzer-HiFi'' refers to \cite{mentzer2020high}, ``Minnen-msssim'' and ``Minnen-mse'' refer to \cite{minnen2018joint}, ``iWave++'' refers to \cite{ma2020end}.}
  \label{RD curves}
\end{figure*}

The results are shown in Fig. \ref{RD curves}. It can be seen that the compared methods perform poorly under at least one metric. For example, ``Mentzer-HiFi'' provides much lower LPIPS, however, this is at the cost of poor PSNR. Differently, our method performs well under both metrics. Compared with the baseline method ``iWave++,'' our method provides a similar PSNR, however, achieves much lower LPIPS. This shows the superiority of the probabilistic image decoding.

\section{Conclusion}
We have proposed an image compression method that allows probabilistic decoding.
Unlike previous image compression methods that choose reconstructed image with maximum likelihood, our method firstly estimates the distribution of all possible images that can generate the same bitstream, and then samples image from the estimated distribution.
By taking different sampling strategies, the decoder could produce diverse images of different characteristics. As an example, we have shown that we can obtain images with high signal fidelity and images with high visual quality based on the same bitstream.

Some modules in the proposed method may be not optimized well. For example, using advanced reversible transform module and probability model, the distribution of possible inputs can be estimated more accurately under the same bit rate. In addition, we only control the sampling process by uniformly adjusting the variances of the estimated distribution. More efficient sampling strategies are worthy of further research.

{\small
\bibliographystyle{plain}

}

\end{document}